\documentclass[10pt,letterpaper]{article}
\usepackage{opex3}
\usepackage{graphicx}% Include figure files
\usepackage{bm}% bold math

\def\beq{\begin{equation}}
\def\eeq{\end{equation}}
\def\beqa{\begin{eqnarray}}
\def\eeqa{\end{eqnarray}}

\newcommand{\roughly}[1]{\mathrel{\raise.3ex\hbox{$#1$\kern-0.85em
\lower1ex\hbox{$\sim$}}}}

\begin{document}

%\preprint{APS/123-QED}

\title{Efficient Terahertz Generation in Triply Resonant Nonlinear Photonic Crystal Microcavities}% Force line breaks with \\

\author{Ian B. Burgess$^{1\dagger}$, Yinan Zhang$^{1\dagger}$, Murray W. McCutcheon$^{1\dagger}$,\\ Alejandro W. Rodriguez$^{2}$, Jorge Bravo-Abad$^{2}$, Steven G. Johnson$^{3}$, Marko Lon{\v{c}}ar$^{1*}$}
\vspace{10pt}
 \email{loncar@seas.harvard.edu}
\address{{\small $^1$ School of Engineering and Applied Sciences, Harvard University, Cambridge, MA 02138}\\
{\small $^2$ Department of Physics, Massachusetts Institute of Technology, Cambridge, MA, 02139}\\
{\small $^3$ Department of Mathematics, Massachusetts Institute of Technology, Cambridge, MA, 02139}\\
{\small $\dagger$} These authors contributed equally.}

\date{March 2009}

\begin{abstract}
We propose a scheme for efficient cavity-enhanced nonlinear THz generation via difference-frequency generation (DFG) processes using a {\it triply} resonant system based on photonic crystal cavities. We show that high nonlinear overlap can be achieved by coupling a THz cavity to a doubly-resonant, dual-polarization near-infrared (e.g. telecom band) photonic-crystal nanobeam cavity, allowing the mixing of three mutually orthogonal fundamental cavity modes through a $\chi^{(2)}$ nonlinearity. We demonstrate through coupled-mode theory that complete depletion of the pump frequency~--~i.e., quantum-limited conversion~--~is possible in an experimentally feasible geometry, with the operating output power at the point of optimal total conversion efficiency adjustable by varying the mode quality ($Q$) factors.
\end{abstract}
%\pacs{42.65.Sf, 42.65.Tg, 42.65.-k}
%\hoffset 20pt
%%%%%%%%%%%%%%%%%%%%%%% References %%%%%%%%%%%%%%%%%%%%%%%%%

%%%%%%%%%%%%%%%%%%%%%%%%%%  body  %%%%%%%%%%%%%%%%%%%%%%%%%%
%\maketitle
\section{Introduction}
Nonlinear optical frequency conversion is widely used for the generation of light in parts of the spectrum for which there are no convenient sources \cite{30}. In particular, nonlinear processes are regarded as a promising route to generation of coherent radiation in the terahertz (THz) frequency range \cite{2,6,18,19,20,23,24,27,28}. Recently, there has been renewed interest in cavity-enhanced nonlinear frequency conversion, as improved designs and fabrication techniques have paved the way for the realization of wavelength-scale cavities, thus allowing efficient conversion at increasingly low powers \cite{2,6,18,19,20,1,4,5,10,11,12,13,25}.
\begin{figure}[t]
\centering
\includegraphics[width=11.0cm]{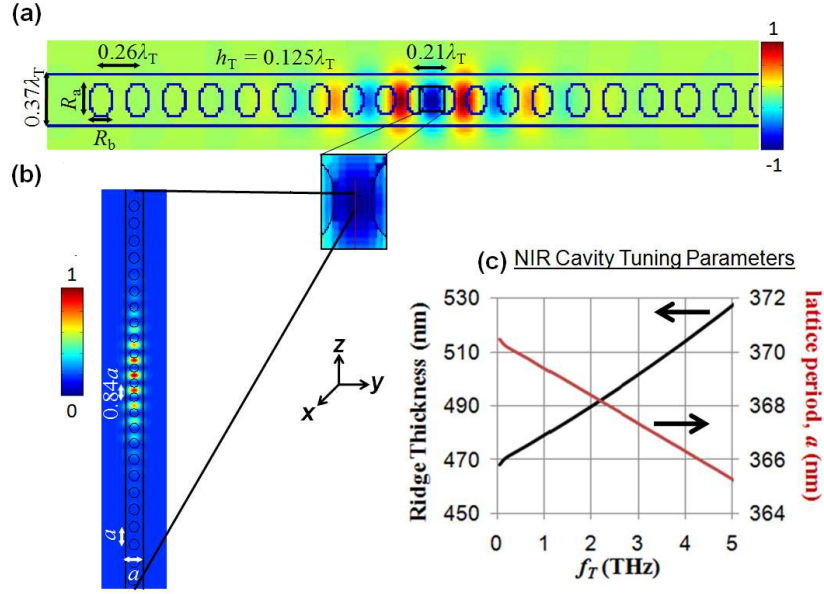}
\caption{\label{T1} Design of the triply resonant system of coupled photonic crystal nanobeam cavities for efficient THz generation. A dual-mode PhC nanobeam cavity is suspended just above the THz cavity (there is a $1\mu$m air gap in our simulations) near its field maximum. {\bf (a)} Normalized mode profile ($E_{z,\mathrm{T}}$) of the THz mode. Dimensions are shown as a function of the THz wavelength. Dimensions are shown here for GaAs, but in principle any material and design can be used as long as it has a THz mode with the correct polarization. {\bf (b)} Normalized nonlinear mode-product ($E_{y,\mathrm{TE}}E_{x,\mathrm{TM}}$)of the TE-like and TM-like modes in the dual-mode cavity which overlaps with the THz mode shown. This product has a single sign, indicating constructive overlap of the two modes. Dimensions other than the beam thickness are shown as a function of the lattice period $a$. {\bf (c)}: Nanobeam thickness (left) and Bragg mirror periodicity, (right) for the NIR PhCNC plotted as a function of the THz difference frequency. The TE-like mode frequency is fixed at 200THz ($\lambda_{\mathrm{TE}}=1.5\mu\mathrm{m}$, $\lambda_{\mathrm{TE}}<\lambda_{\mathrm{TM}}$).}
\end{figure}

In this paper, we propose a scheme for efficient THz generation based on second-order $\chi^{(2)}$ difference frequency process in a triply resonant structure consisting of two nested photonic-crystal cavities, as shown schematically in Fig.~1. The first cavity, designed to operate at near-infrared (NIR) wavelengths and realized in a nonlinear material (e.g. III--V semiconductors), produces nonlinear-polarization at THz frequencies by difference-frequency generation (DFG). The cavity is placed in the proximity of a second, much larger, single-mode cavity with the resonance in the THz range. We utilize the fact that the THz wavelength, $\lambda_{\mathrm{T}}$ ($\sim150\mu$m), is on a vastly different length scale from the pump and idler wavelengths, $\lambda_{1}$ and $\lambda_{2}$ respectively ($\sim1.5\mu$m), allowing us to effectively decouple the cavity designs. Our system can be viewed as a classical version of a two-level atom strongly coupled to the optical cavity. In our design, we suspend the telecom band cavity above the THz cavity near the field maximum \cite{31}, sufficiently far so that the telecom modes are not affected by the presence of the THz cavity, but sufficiently close that the long-wavelength THz mode extends substantially into this cavity (e.g. 1$\mu\mathrm{m}$ above). In this way we can achieve a high overlap between the three {\it fundamental} cavity modes by mixing three mutually orthogonal polarizations through the strong $\chi^{(2)}_{ijk}$ ($i\neq j\neq k$) terms in the nonlinear susceptibility of III--V semiconductors (e.g. GaAs, GaP) \cite{2,21}. This scheme takes advantage of our recently demonstrated doubly-resonant ultrahigh-$Q$ photonic-crystal nanobeam cavities (PhCNC) with mutually orthogonally polarized modes (i.e. one is TE-like one is TM-like) \cite{27,8}. Both cavities are preferentially coupled to a corresponding waveguide extending from one end, with the length of the Bragg mirror of holes at one end used to tune the strength of the coupling. The mirror at the other end is made sufficiently long such that leakage in that direction is much smaller than out-of-plane losses. Since the relevant nonlinear material is in the NIR cavity, the THz cavity can be composed of any material, a useful degree of freedom considering the scarcity of low-loss THz materials.

\section{Model}
We employ the coupled-mode theory (CMT) framework described in Ref.~\cite{10} to analyze triply resonant THz generation by DFG in a wavelength-scale cavity with a non-resonant $\chi^{(2)}$ nonlinear susceptibility. Our cavity has three resonant modes at frequencies $\omega_{1}$ (pump), $\omega_{2}$ (idler), and $\omega_{\mathrm{T}}$ (THz signal), which satisfy $\omega_{\mathrm{T}}=\omega_{1}-\omega_{2}$ and each mode is coupled to a waveguide. As described in Ref.~\cite{10}, the conversion efficiency of the system is determined by the input powers, ($P_{1},P_{2}$), the cavity mode quality factors, {$Q_{k}$} ($k=1,2,\mathrm{T}$), the cavity waveguide coupling strengths {$\Gamma_{k}$} (defined as $\Gamma_{k}\equiv\gamma_{k,s}/\gamma_{k}$, where $\gamma_{k,s}$ is the leakage rate into the waveguide and $\gamma_{k}$ is the total leakage rate including linear losses), and the nonlinear coupling constant $\beta$, defined in Ref.~\cite{10}. As discussed in Ref.~\cite{5}, avoiding mode symmetries and polarizations that lead to zero coupling constant, as well as optimizing the value of $\beta$, is analogous to phase-matching in the nonlinear mixing of propagating modes. For our type of design, since the THz field strength is nearly constant across the much smaller NIR cavity, $\beta$ is well approximated by:
\beq
\beta\approx\frac{\kappa_{\mathrm{T}}d_{\mathrm{eff}}}{2\sqrt{\epsilon_{0}\lambda_{\mathrm{T}}^{3}}}\frac{\int_{\mathrm{d}} d^3{\bf r}E_{\mathrm{TE},y}^*E_{\mathrm{TM},x}}{\sqrt{\int d^3{\bf r}\epsilon_{r}|{\bf E}_{\mathrm{TE}}|^2}\sqrt{\int d^3{\bf r}\epsilon_{r}|{\bf E}_{\mathrm{TM}}|^2}}\,,\label{E2}
\eeq
where ${\bf E_{\mathrm{TE}}}$, ${\bf E_{\mathrm{TM}}}$, and ${\bf E_{\mathrm{T}}}$ are the field profiles of the cavity modes, $\int_{\mathrm{d}}$ denotes spatial integration only over the nonlinear dielectric, and $\kappa_{\mathrm{T}}$ is a dimensionless constant that quantifies the THz cavity contribution to the nonlinear overlap, defined as: $\kappa_{\mathrm{T}}\equiv\lambda_{T}^{3/2}E_{z,\mathrm{T},\mathrm{NIR}}/[\sqrt{\int d^3{\bf r}\epsilon_{r}|{\bf E}_{\mathrm{T}}|^2}]$, where $E_{z,\mathrm{T},\mathrm{NIR}}$ is the constant THz field amplitude inside of the NIR cavity.

As described in Ref.~\cite{10}, quantum conversion efficiency is quantified by the parameter:
\beq
E_{\mathrm{ff}}^{\mathrm{Q}}=\frac{\omega_{1}}{\omega_{\mathrm{T}}\Gamma_{1}\Gamma_{\mathrm{T}}}\frac{P_{\mathrm{out},\mathrm{T}}}{P_{1}}\,.\label{E2}
\eeq
This parameter describes the number of THz photons collected per photon of signal ($\omega_{1}$) input that couples into the cavity (factoring in losses). As expected, this parameter is bounded in the steady state by $E_{\mathrm{ff}}^{\mathrm{Q}}\le1$ \cite{10}. Since the power efficiency ($P_{\mathrm{out},\mathrm{T}}/P_{1}$) is scaled by a factor of $1/\Gamma_{1}\Gamma_{\mathrm{T}}$, a decrease in $\Gamma_{1}\Gamma_{\mathrm{T}}$ causes a decrease in the maximum attainable power efficiency (always occurring at $E_{\mathrm{ff}}^{\mathrm{Q}}=1$). Therefore over-coupling of the cavity modes to the corresponding input/output waveguides is imperative for optimal performance~\cite{10}. As derived in Ref.~\cite{10}, the efficiency of conversion for any triply resonant DFG system can be expressed as a function only of the input powers, normalized against the critical powers ($P_{1,\mathrm{crit}}$, $P_{2,\mathrm{crit}}$) defined as,
\beq
P_{k,\mathrm{crit}}\equiv\frac{\omega_{k}}{16\tilde{Q}\Gamma_{k}|\beta|^2}\,,\label{E3}
\eeq
where $\tilde{Q}\equiv Q_{1}Q_{2}Q_{\mathrm{T}}$. This means that only the product of the three $Q$ factors contributes to efficiency and the distribution of this product is unimportant. Figure~2a shows the stable quantum efficiencies reached after a simple continuous-wave (CW) step excitation as a function of normalized input powers. The dotted black line denotes the onset of bistability to its right~\cite{10}. Quantum-limited conversion ($E_{\mathrm{ff}}^{\mathrm{Q}}=1$) can be achieved when the input powers satisfy $P_{2}/P_{2,\mathrm{crit}}=(1-P_{1}/4P_{1,\mathrm{crit}})^2$. We can also define a total-efficiency parameter that is proportional to the ratio of THz output power to total input power at both telecom frequencies:
\beq
E_{\mathrm{ff}}^{\mathrm{Tot}}\equiv\frac{\omega_{1}}{\omega_{\mathrm{T}}\Gamma_{1}\Gamma_{\mathrm{T}}}\frac{P_{\mathrm{out},\mathrm{T}}}{P_{1}+P_{2}}=\frac{E_{\mathrm{ff}}^{\mathrm{Q}}P_{1}}{P_{1}+P_{2}}\,.\label{E5}
\eeq
Quantum efficiency and total efficiency are equivalent in the asymptotic limit, $P_{2}\to0$ ($P_{2}\ll P_{2,\mathrm{crit}}$, $P_{2}\ll P_{1}$). In this limit, the conversion efficiency is mono-stable~\cite{10}. Total power conversion is optimized when $P_{2}\ll P_{2,\mathrm{crit}}$, $P_{1}=4P_{1,\mathrm{crit}}$. One can calculate a trajectory in the ($P_{1}$,$P_{2}$) space that maximizes $E_{\mathrm{ff}}^{\mathrm{Tot}}$. This corresponds to the optimal operating conditions for a single cavity (fixed $\tilde{Q}$, \{$\omega_{k}$\}, $\beta$) operating through a range of powers. The white dashed curve in Fig.~2a shows the ideal operating conditions for CW THz generation (the limit $\omega_{1}\approx\omega_{2}$). For $P_{1}>4P_{1,\mathrm{crit}}$, there do exist solutions with slightly higher efficiency than on this trajectory, but these solutions are only stable for certain $Q$-factor combinations and cannot be excited with a step CW excitation. Keeping $P_{2}<<P_{1}$ for all $P_{1}>4P_{1,\mathrm{crit}}$ ensures mono-stable behavior in this region~\cite{10}.
\begin{figure*}[top]
\centering
\includegraphics[width=13.5cm]{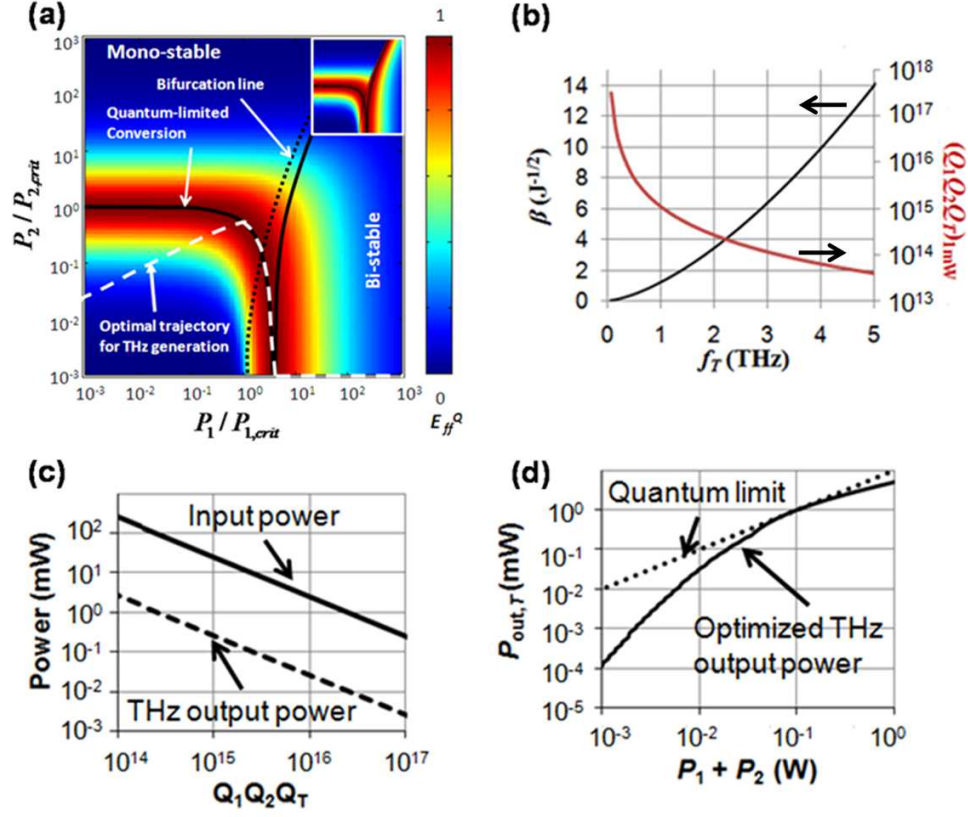}
\caption{\label{T3} {\bf (a)} Step-excitable quantum efficiency ($E_{\mathrm{ff}}^{\mathrm{Q}}$) for stable CW THz generation,  plotted as a function of powers of the pump and idler normalized against the critical powers $P_{k,\mathrm{crit}}$ (see Eq.~(\ref{E3})). The solid line denotes the critical relationship between input powers where $E_{\mathrm{ff}}^{\mathrm{Q}}=1$ is possible ($P_{1}=4P_{1,\mathrm{crit}}$). The black dotted line denotes the onset of bistability. An inset showing the other stable solution is shown in the top right corner~\cite{10}. The white dashed line shows the optimal operating conditions for maximum total conversion efficiency {\bf (b)}: Performance parameters of our nested PhCNC design (GaAs) as a function of the THz resonance frequency: nonlinear overlap, $\beta$ (left), and the $Q$-factor product required for 1mW of THz power to be generated from a pump at 200THz with quantum limited efficiency (right). {\bf (c)}: Dependence of the input power (solid line) yielding optimal efficiency ($P_{1} = 4P_{1,\mathrm{crit}}$) and the corresponding THz output power (dashed line) on the cavity $Q$-factor product ($\tilde{Q}$), for coupled GaAs THz and dual mode NIR PhCNCs ($\omega_{1}/2\pi$, $\omega_{2}/2\pi\sim200$THz, $\omega_{\mathrm{T}}/2\pi\sim2.0$THz, $\beta\sim3.5J^{-1/2}$). {\bf (d)}: THz output power as a function of input power in this geometry for, $\tilde{Q}\sim2.5x10^{14}$. The dotted line shows the quantum limit.}
\end{figure*}

\section{Design}
In order to estimate the coupling constant, $\beta$, we consider a geometry similar to that shown in Fig.~1: a PhCNC, with closely-spaced TE-like and TM-like modes in the near infrared (NIR) telecom spectral region ($\omega_{1}/2\pi$, $\omega2/2\pi\sim200$THz)~ \cite{8} positioned 1$\mu$m above a larger THz cavity. Our cavities, both telecom and THz, are based on a free-standing ridge waveguide patterned with a one-dimensional (1D) lattice of holes, utilizing a tapered photonic-crystal cavity approach that we and others have recently demonstrated~\cite{27,9,17,26,32}. We consider the case where both cavities are composed of GaAs ($n_{\mathrm{NIR}}\approx3.45$, $n_{\mathrm{THz}}\approx3.6$)~\cite{28,21}, and the telecom cavity is suspended slightly above the THz cavity at a height, $d=1\mu\mathrm{m}\ll\lambda_{\mathrm{T}}$. By engineering the "impedance matching" between the cavity mode and evanescent Bloch mode that exists in the photonic crystal (Bragg) mirror we have demonstrated, both experimentally and theoretically, $Q$ factors on the order of $10^6$~\cite{27,17}. A 3D finite-difference time-domain code (3D-FDTD)~\cite{29} was used to design independent telecom and THz cavities~\cite{33}, and a CMT model (as described above) was used to describe the nonlinear frequency conversion.

We use the design algorithm detailed in Ref.~\cite{27} for the design of an NIR PhCNC supporting orthogonal TE-like and TM-like modes, which is necessary to achieve efficient nonlinear conversion as dictated by the $\chi^{(2)}$ tensor of a III--V semiconductor material. We fix the ridge width equal to the regular lattice period ($a$) and use circular holes in the Bragg-mirror region with radius, $R=0.3a$. The period and hole radius are then tapered linearly in reciprocal space to 0.84 times the normal value over 8 periods~\cite{27}. All cavity dimensions are fixed with respect to the Bragg-mirror lattice period ($a$) except for the ridge thickness (or height), which is used to tune the frequency spacing between the TE-like and TM-like modes (see Fig.~1d and Ref.~\cite{27}). Here we vary the cavity dimensions so that the TE-like resonance is always at 200THz (1.5$\mu$m) and the TM-like resonance is at a slightly lower frequency, giving the desired THz difference frequency. The TM-like resonance can also be tuned above the TE-like resonance in frequency, but we find that this gives poorer TM $Q$ factors and lower values of $\beta$ as the TM-like bandgap shrinks in this direction~\cite{27}. The NIR cavity parameters are plotted in Fig.~1c as a function of the desired THz frequency.

For the THz mode, we consider a TE-like mode of a second GaAs PhCNC, based on the design shown in Fig.~1a, having elliptical holes ($R_{\mathrm{a}}=1.5R_{\mathrm{b}}$) and a four-hole, linear taper of the lattice period and hole diameters as was done with the NIR cavity. Fig.~1a shows the cavity dimensions as a function of the free-space wavelength of the resonance. This cavity can be scaled to fit the desired THz frequency and has $\kappa_{\mathrm{T}}=0.65$ for any scaling. This cavity could in principle be replaced by any type of THz cavity having the same value of $\kappa_{\mathrm{T}}$, and all operating properties described below would be unchanged. This degree of freedom is a key advantage of our decoupled-design paradigm. The power at which maximum conversion efficiency occurs (normalized $P_{1}=4P_{1,\mathrm{crit}}$) for a given THz frequency depends on the product of the three $Q$ factors and the overlap, $\beta$. Fig.~2b shows the dependence on the THz frequency in our design scheme of $\beta$ and the $Q$-product required to produce 1mW of THz radiation with optimal efficiency. Figure~2c shows the optimized NIR input power and optimized THz output power as a function of the $Q$-product for our system for $f_{\mathrm{T}}=2$THz, $\beta=3.5\mathrm{J}^{-1/2}$.

The $Q$~factors in our cavities were originally designed to be limited by out-of-plane scattering~\cite{27,9}. Overcoupling to the input/output waveguide is then achieved by reducing the length of the Bragg mirror on one side. This reduces the maximum values of $Q$ by an order of magnitude, but minimizes the reduction in efficiency caused by losses~\cite{10}, and enables operation close to quantum-limited conversion. We find that this variation of the Bragg mirror lengths has negligible effect on the overlap $\beta$ for an order of magnitude variation in the $Q$-factors. For some applications where a single cavity must operate across a range of different power levels, the best performance is achieved by varying the two input powers to fit the trajectory described in the previous section (see Fig.~2a). For example, this optimal output power is shown in Fig.~2d for a cavity optimized to produce 1mW of power at 2 THz. This optimization requires a $Q$-factor product of $\tilde{Q}=2.5\times10^{14}$. When limited by out-of-plane losses, the $Q$ factors of our TE/TM dual-mode NIR cavity are $Q_{1}(\mathrm{TE})=2\times10^7$, $Q_{2}(\mathrm{TM})=3\times10^6$ for $f_{\mathrm{T}}=2$THz. Reduction of each mode's $Q$ factor by an order of magnitude to overcouple the waveguide requires a THz $Q$ of only $\sim300$. This low requirement is important since realistic THz cavities will be limited by material losses in the THz range. While GaAs has a low absorption coefficient compared to most materials in the THz range, its linear loss rate still corresponds to a $Q$ factor of about $1.5\times10^3$~\cite{28}, which is well below the scattering-limited $Q$ of our design ($Q_{\mathrm{T},\mathrm{scat}}=1.4\times10^6$). THz losses are likely to put an upper bound on any THz cavity $Q$ factor in practice, and further reduction of this limit is then required for overcoupling. However, a large advantage of this geometry is that all nonlinear processes take place outside of the THz cavity material in the NIR cavity, so that any design and any material can be used for the THz cavity, with the performance of any THz cavity coupled to our dual-mode nanobeam cavity purely determined by the cavity's $\kappa_{\mathrm{T}}$ value. Thus, this loss limit can be increased with the discovery of more efficient materials in the THz wavelengths.
\begin{figure}[top]
\centering
\includegraphics[width=8.5cm]{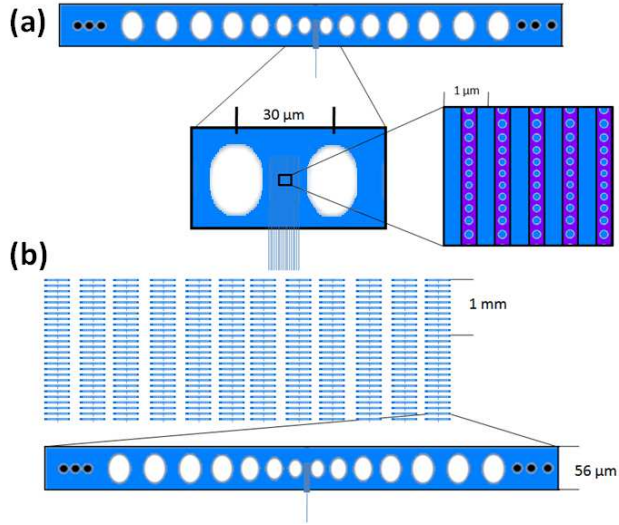}
\caption{\label{T4} {\bf (a)}: Large phased array of $N$ NIR cavities placed on a single THz cavity. This array will have the same THz- generation performance properties as the corresponding device having only a single NIR cavity, however the damage threshold input power will increase by a factor of $N$. {\bf (b)}: High-power tunable generation is possible by incorporating many ($>100$) THz devices on a single chip. An array of $N$ devices that emit $P_{\mathrm{out},\mathrm{T}}$ most efficiently allows quantum limited conversion over the full range of output powers from $P_{\mathrm{out},\mathrm{T}}$ to $NP_{\mathrm{out},\mathrm{T}}$. The specific scale shown in both (a) and (b) is for $\lambda_{\mathrm{T}}=150\mu$m.}
\end{figure}

\section{Arrays}
Considering that high-$Q$ NIR modes are needed to completely convert even relatively high pump powers (see Fig.~2), the material damage threshold is likely to become an important design parameter. The vast difference in length-scales in our design can be further exploited to circumvent this problem. This is because the NIR cavity is so small compared to the THz mode profile that many NIR cavities can be fabricated on top of a single THz cavity, all within the fundamental anti-node. In this limit, where the NIR cavity occupies a very small volume of the THz mode and where there is good overlap between the NIR modes, the NIR mode volumes have a negligible effect on $\beta$ [see Eq.~(\ref{E2})]. Therefore, we can consider an array of $N$ NIR cavities on the anti-node of the THz cavity (whose inputs are phase-locked), shown in Fig.~3a, as a single NIR cavity. This cavity has the same $Q$ factor and a $N$ times higher mode volume, but will have the same $\beta/\kappa_{\mathrm{T}}$, yielding an unchanged value of $\beta$ provided that the array still fits entirely inside a region of the fundamental anti-node with roughly constant THz field amplitude. Thus, this cavity will have the same optimal THz output power, but with a damage threshold that is increased $N$-fold.

Quantum-limited efficient conversion can be achieved over a broad range of powers from a single chip by integrating many triply resonant devices as shown in Fig.~3b. The total size of a single THz device is given by the size of the THz cavity, which can be sub-millimeter (i.e., on the THz wavelength scale). Thus, hundreds of these devices can be integrated on a single square-centimeter chip. Large arrays allow not only for high-power efficient generation, but also maximally efficient generation over a broadly tunable range compared to what can be achieved from a single device. To see this, suppose a single device emits $P_{\mathrm{out},\mathrm{T}}$ of THz power with $E_{\mathrm{ff}}^{\mathrm{Tot}}=1$. An array of $N$ of these devices therefore can emit $NP_{\mathrm{out},\mathrm{T}}$ with $E_{\mathrm{ff}}^{\mathrm{Tot}}=1$, but can also emit at this efficiency over the whole range $P_{\mathrm{out},\mathrm{T}}\geq~P\leq~ NP_{\mathrm{out},\mathrm{T}}$ by varying the number of devices in the array that are used.

\section{Conclusion}
We have proposed an experimentally feasible platform for high efficiency THz generation in triply resonant photonic crystal cavities. By placing a dual-mode photonic-crystal nanobeam cavity with closely spaced TE-like and TM-like resonant modes in the telecom band near the field maximum of a much larger THz cavity mode, we have shown that a high nonlinear overlap of mutually orthogonal modes can be attained in many nonlinear materials (eg III--V semiconductors). Total power conversion approaches the quantum limit at a critical power which can be tuned by varying the product of the three mode $Q$ factors. Coupling of many NIR cavities to a single THz cavity increases the material damage threshold without changing the device performance properties. High-efficiency power output can be further increased and tuned by fabricating large arrays of these triply resonant devices on a single chip.

\section{Acknowledgements}
We would like to thank J.D. Joannopoulos and Marin Solja{\v{c}}i{\'{c}} for helpful discussions. IBB and MWM wish to acknowledge NSERC (Canada) for support from PGS and PDF programs. This work is supported through NSEC at Harvard, by the Army Research Office through the ISN under Contract No. W911NF-07-D-0004, and by US DOE Grant No. DE-FG02-97ER25308 (ARW).


\begin{thebibliography}{99}
\bibitem{31} At the position of highest THz field that exists above the THz material, the field amplitude has $\sim25\%$ of the maximum field amplitude for our THz nanobeam design.
\bibitem{33} The NIR cavity was included in the THz cavity simulations and it was found to cause a slight decrease in the scattering-limited $Q$ factor ($2\times10^6\rightarrow1.4\times10^6$), while having a negligible effect on $\kappa_{\mathrm{T}}$.
\end{thebibliography}
\end{document}